\documentclass[pra,aps,twocolumn,eqsecnum,superscriptaddress,showpacs]{revtex4}

\usepackage{amsmath,amssymb}
\usepackage{graphicx}
\usepackage{color}
\usepackage{hyperref}
\hypersetup{colorlinks=true,citecolor={blue},linkcolor={blue},urlcolor={blue},pdfstartview=FitH}
\usepackage{ulem}

\begin{document}

\title{Light Scattering From an Atomic Array Trapped Near \\ a One-Dimensional Nanoscale Waveguide: a Microscopic Approach}

\author{V.A. Pivovarov}
\affiliation{Physics Department, St.-Petersburg Academic University, Khlopina 8, 194021 St.-Petersburg, Russia}
\author{A.S. Sheremet}
\affiliation{Laboratoire Kastler Brossel, UPMC-Sorbonne Universit\'es, CNRS, ENS-PSL Research University, Coll\`ege de France, 4 place
Jussieu, 75005 Paris, France}
\affiliation{Russian Quantum Center, Novaya 100, 143025 Skolkovo, Moscow Region, Russia}
\author{L.V. Gerasimov}
\affiliation{Faculty of Physics, M.V. Lomonosov Moscow State University, Leninskiye Gory 1-2, 119991, Moscow, Russia}
\author{V.M.\nolinebreak\ Porozova}
\affiliation{Department of Theoretical Physics, St-Petersburg State
Polytechnic University, 195251, St.-Petersburg, Russia}
\author{N.V. Corzo}
\affiliation{Laboratoire Kastler Brossel, UPMC-Sorbonne Universit\'es, CNRS, ENS-PSL Research University, Coll\`ege de France, 4 place
Jussieu, 75005 Paris, France}
\author{J. Laurat}
\affiliation{Laboratoire Kastler Brossel, UPMC-Sorbonne Universit\'es, CNRS, ENS-PSL Research University, Coll\`ege de France, 4 place
Jussieu, 75005 Paris, France}
\author{D.V. Kupriyanov}
\email{kupr@dk11578.spb.edu}
\affiliation{Department of Theoretical Physics, St-Petersburg State
Polytechnic University, 195251, St.-Petersburg, Russia}

\date{\today}

\begin{abstract}
\noindent The coupling of atomic arrays and one-dimensional subwavelength waveguides gives rise to interesting photon transport properties, such as recent experimental demonstrations of large Bragg reflection and paves the way for a variety of potential applications in the field of quantum non-linear optics. Here, we present a theoretical analysis for the process of single-photon scattering in this configuration using a full microscopic approach. Based on this formalism, we analyze the spectral dependencies for different scattering channels from either ordered or disordered arrays. The developed approach is entirely applicable for a single-photon scattering from a quasi-one-dimensional array of multilevel atoms with degenerate ground state energy structure. Our approach provides an important framework for including not only Rayleigh but also Raman channels in the microscopic description of the cooperative scattering process.
\end{abstract}


\maketitle

\section{Introduction}

\noindent  Efficient control of light-matter interaction at the single-photon level is a central and challenging task for quantum optics and quantum information science \cite{Kimble2008, DLCZ2001,Chou2007,reviewensemble,Chang2007, Chang2014,Kupriyanov17}. At the fundamental level this interaction manifests itself via the basic quantum electrodynamics processes of spontaneous emission, absorption and scattering of a single photon by a single atom. In free space, the efficiency of this interaction is limited by the small atomic scattering cross section in comparison to usual large beam shining area. However, it can be greatly enhanced by placing a single atom in the vicinity of subwavelength waveguide due to the Purcell effect \cite{Purcell1946} or by considering large atomic ensembles \cite{Dicke1954}. The combination of these approaches has motivated recent experimental efforts towards the development of novel platforms that integrate an atomic chain coupled with a dielectric nanoscale waveguide \cite{Kimble_cavity2012,Ritsch2012a,Chang2013,Kimble2014,Goban2015,Review2016,Review2017}.

One example of such platforms is the so-called subwavelength nanofiber \cite{Klimov2004,Hakuta2005,Hakuta2005_2atoms,Hakuta2008PRA,Review2016,Review2017}. Due to a large evanescent field, the guided light can be efficiently used for trapping atoms close to the surface \cite{Balykin2004,Rauschenbeutel2010,Kimble_trapping2012} and interacting with them\cite{Hakuta2007,Hakuta2008NJP,Ritsch2012b,Hakuta2012}. Recent experiments include for instance the demonstration of optical memories in this setting \cite{Gouraud2015,Syrin2015}. The trapping technique can also be adjusted for arranging atoms in an optical lattice commensurate or nearly-commensurate with the resonant wavelength. Such capability enables to investigate Bragg reflection \cite{Corzo2016,Polzik16} and long-range interactions \cite{Rolson2017}.

These recent experiments and supporting theoretical studies \cite{LeKien_propagation,LeKien_anysotropy,Kornovan2016,Asenjo-Garcia2017} have shown that light scattering from atoms interacting with the evanescent field of the waveguide mode has important differences from the light scattering from atoms in free space. The effects of Zeeman degeneracy of atomic transition should be taken into consideration for relevant description of the Raman channels, which at present was primary studied in free space \cite{Sheremet12,Sheremet15}. The problem becomes more complicated as the strong field confinement provided by the nanofiber imposes an inherent link between the local polarization and propagation direction of light. In such strongly non-paraxial regime, the spin and orbital angular momentum of the guided light obey joint dynamics and cannot be independently considered. In particular it leads to direction-sensitive emission and absorption of the guided light by the atoms \cite{Petersen2014,Mitsch2014,Lodahl2015,Lodahl2017,Corzo2016}.

All these experimental results and theoretical works demonstrate that such a hybrid system is a versatile platform for studying various cooperative effects emerging from both the collective atomic structuring and the waveguide-mediated long-range coupling between atoms. The internal correlations, existing in a many particle microscopic quantum entangled state, can play an essential role in the cooperative scattering process. In this context, an ab-initio theoretical insight on the problem of light-atoms interaction would provide a natural extension of the convenient but sometimes insufficient self-consistent description, based on the density matrix approach and the Maxwell-Bloch formalism. The present paper develops such a systematic microscopic theory of light scattering from an atomic array trapped near a nanofiber with taking into account the entire interaction dynamics together with the complete angular momentum structure of the guided light and the degenerate energy structure of the atoms.

The paper is organized as follows. In Section \ref{Section_II} we first provide a general description of our microscopic approach of light scattering in a waveguide configuration and emphasize the differences with similar process in free space. The derivation details concerning the modification of the electric field Green's function near the nanofiber are given in Appendix \ref{Appendix_A}.  In Section \ref{Section_III} we then present numerical simulations for light scattering from an array consisting of five atoms. Ordered and disordered configurations are considered. The parameters of the waveguide mode used in the simulation are given in Appendix \ref{Appendix_B}. Section \ref{Section_IV} finally concludes the paper.

\section{The scattering problem revised for a waveguide configuration}\label{Section_II}

\noindent In this section, we review the basic points of the quantum scattering problem. Normally introduced in terms of transformation of plane waves, it is extended here to the specific configuration where the input and output states are fundamental modes of the waveguide.

\subsection{Mathematical framework}

\noindent The mathematical framework of the quantum scattering problem is based on the concept of the asymptotic evolutionary operator $\hat{S}$ that transforms the system states from infinite past $|\psi\rangle_{\mathrm{in}}$ to infinite future $|\psi\rangle_{\mathrm{out}}$ as a result of the interaction process \cite{GoldbergerWatson}. In the interaction representation, the corresponding asymptotic transformation is given by
\begin{equation}
|\psi\rangle_{\mathrm{out}}=\mathrm{e}^{\frac{i}{2\hbar}H_0\tau}\mathrm{e}^{-\frac{i}{\hbar}H\,\tau}\mathrm{e}^{\frac{i}{2\hbar}H_0\tau}|\psi\rangle_{\mathrm{in}}\equiv%
\hat{S}|\psi\rangle_{\mathrm{in}}
\label{2.1}
\end{equation}
where $\tau\to+\infty$. The operator $\hat{S}$ can be represented as a matrix in a decoupled basis of two interacting subsystems, which we specify as $|\phi_i\rangle$ for the initial and $|\phi_f\rangle$ for the final system states
\begin{eqnarray}
S_{fi}&=&\delta_{fi}-2i\,\frac{\sin[(E_f-E_i)\tau/2\hbar]}{E_f-E_i}\,T_{fi}(E_i+i0)%
\nonumber\\%
&\Rightarrow &\delta_{fi}-2\pi i\,\delta(E_f-E_i)\,T_{fi}(E_i+i0)%.
\label{2.2}
\end{eqnarray}
The first term selects non-interactive contribution and the interaction part constructs the $T$-matrix i.e. the transition amplitude between the states with same energy $E_f=E_i$. The latter requires infinitely long "interaction" time $\tau\to +\infty$. The $T$-matrix, considered as function of arbitrary complex energy argument $E$, is expressed as follows
\begin{equation}
\hat{T}(E)=\hat{V}+\hat{V}\frac{1}{E-\hat{H}}\hat{V}%
\label{2.3}
\end{equation}
where
\begin{equation}
\hat{H}=\hat{H}_0+\hat{V}.%
\label{2.4}
\end{equation}
$\hat{H}_0$, $\hat{V}$ and $\hat{H}$ are the unperturbed Hamiltonian, the interacting part and the total system Hamiltonian respectively.

For a particular case of a single-photon scattering on an ensemble of $N$ atomic dipoles the interaction part is given by
\begin{eqnarray}
\hat{V}&=&-\sum_{a=1}^{N}%
\hat{\mathbf{d}}^{(a)}\hat{\mathbf{E}}(\mathbf{r}_a)+\hat{H}_{\mathrm{self}},%
\label{2.5}%
\end{eqnarray}
The first term accumulates partial interactions for each of an $a$-th atomic dipole $\mathbf{d}^{(a)}$ with an electric field $\hat{\mathbf{E}}(\mathbf{r})$ at the point of dipole's position. The second term is the dipoles' self-energy part, which is mainly important for renormalization of the self-action divergencies \cite{ChTnDRGr,SKKH2009,Kupriyanov17}. The field operator can be expressed by a standard expansion in the basis of plane waves as
\begin{eqnarray}
\lefteqn{\hat{\mathbf{E}}(\mathbf{r})\equiv \hat{\mathbf{E}}^{(+)}(\mathbf{r})%
+\hat{\mathbf{E}}^{(-)}(\mathbf{r})}%
\nonumber\\%
&&=\sum_{j}\left(\frac{2\pi\hbar\omega_j}{{\cal V}}\right)^{1/2}\!%
i\left[\mathbf{e}_j a_j\mathrm{e}^{i\mathbf{k}\cdot\mathbf{r}}%
-\mathbf{e}_j^{\ast} a_j^{\dagger}\mathrm{e}^{-i\mathbf{k}\cdot\mathbf{r}}\right]%
\label{2.6}%
\end{eqnarray}%
where $j\equiv{\mathbf{k},\alpha}$ is the complex mode index with the mode wave vector $\mathbf{k}$, frequency $\omega_j=c\,k$ and $\alpha=1,2$ numerating two orthogonal transverse polarization vectors $\mathbf{e}_j\equiv\mathbf{e}_{\mathbf{k}\alpha}$ for each $\mathbf{k}$. Here $a_j$ and $a_j^{\dagger}$ are the annihilation and creation operators for the $j$-th field's mode in free space and the quantization scheme includes the periodic boundary conditions in the quantization volume ${\cal V}={\cal L}^3$.

According to the standard assumptions of the time-dependent scattering theory in free space the infinite "interaction" time $\tau$ is physically associated with the duration with which the photon's wave packet overlaps and passes the scattering object. As far as the wave packet tends to approach the monochromatic state its duration should approach infinity and as a consequence this justifies the presence of $\delta$-function and energy conservation in Eq.(\ref{2.2}). However the transfer $\tau\to+\infty$ should be done only in the final step of the derivation procedure together with constructing the main physical characteristics of the process such as transition probability per unit time, scattering tensor, and cross-section. The derivation is based on the arguments of time-dependent perturbation theory, where the field subsystem is considered as a plane-wave contribution into the states $|\phi_i\rangle$ and $|\phi_f\rangle$ posed with periodic boundary conditions into a certain quantization volume with a length-scale ${\cal L}$ same as assumed in expansion (\ref{2.6}). Because of it, for any wave packet its longest duration $\tau$ is naturally limited by the time-scale associated with the quantization length ${\cal L}$ and both these parameters should approach infinity consistently such that $\tau={\cal L}/c\to +\infty$.

\subsection{Light scattering within waveguide}\label{II.B}

\noindent We now consider the case where the incident photon impinges on a cylindrical single-mode nanoscale dielectric waveguide in its fundamental HE$_{11}$-mode \cite{Marcuse}. We focus on a specific process where both the incident and the scattered photons belong to the fundamental mode, i.e. only forward or backward scattering occur inside the waveguide. Actually there is a continuum of the waveguide modes propagating along the fiber as translational waves parameterized by a longitudinal wave number $k$ and posed with a periodic boundary conditions into a longitudinal quantization segment of length ${\cal L}$. The geometry of the scattering process is shown in Fig.~\ref{fig1}.

For such a specific scattering geometry Eq.(\ref{2.2}) is valid since it is based on a general perturbation theory analysis applied to any quantum system with continuous spectrum. However, the scattering channel is now described by transmission and reflection probabilities instead of scattering cross-section. The singularity of the second term in the definition of the $S$-matrix can be naturally regularized by associating the initial and final states with the wave packets having longitudinal length ${\cal L}$. In accordance with the physical interpretation of time $\tau$ as the duration with which the photon's wave packets overlap and pass the scattering object we have $\tau={\cal L}/v_g\to \infty$ where $v_g=d\omega/dk$ is the group velocity for the fundamental mode at frequency $\omega$ associated with either incident or scattered photons. We neglect here possible difference in group velocities for the case of Raman-type scattering caused by transitions in atomic spin subsystem.

Thus for the scattering within waveguide modes and in the limit ${\cal L}\to\infty$ the singular form of relation (\ref{2.2}) can be regularized as follows
\begin{eqnarray}
S_{i'i}&=&\delta_{i'i}-i\,\frac{{\cal L}}{\hbar v_g}\,T_{i'i}(E_i+i0)%
\label{2.7}
\end{eqnarray}
where we redefined $f=i'$ emphasizing thereby the physical equivalence of initial and final states for light scattering within the waveguide modes. The energy conservation $E_{i'}=E_i$ is fulfilled in the applied regularization and we shall further treat the $S$-matrix components given by Eq.(\ref{2.7}) in its original physical meaning as probability amplitudes between the initial and final states.

As a next step we keep the interaction operator in form of Eq.(\ref{2.5}) but re-expand the field operator (\ref{2.6}) in the complete basis of the guided and external modes
\begin{equation}
\hat{\mathbf{E}}(\mathbf{r})=\sum_{s}\left(2\pi\hbar\omega_s\right)^{1/2}\!%
i\left[b_s\mathbf{E}^{(s)}(\mathbf{r})%
- d_s^{\dagger}\mathbf{D}^{(s)\ast}(\mathbf{r})\right]\ + \ldots%
\label{2.8}%
\end{equation}%
Here we selected only contribution of the guided modes, enumerated by the mode index $s$, for the electric field $\mathbf{E}^{(s)}(\mathbf{r})$ and the displacement field $\mathbf{D}^{(s)}(\mathbf{r})=\epsilon(\mathbf{r})\,\mathbf{E}^{(s)}(\mathbf{r})$, where $\epsilon(\mathbf{r})$ is the spatially dependent dielectric permittivity of the entire medium (free space and the dielectric nanofiber). The modes are specified by Eqs.~(\ref{a.6}) and (\ref{a.8}) in Appendix \ref{Appendix_A}, and the ellipsis in Eq.~(\ref{2.8}) denote the contribution of external modes. The mode operators are given by
\begin{eqnarray}
b_s\!&=&\!\sum_{\mathbf{k},\alpha=1,2}\!\!a_{\mathbf{k}\alpha} \left(\frac{\omega_k^{\mathrm{free}}}{\omega_s^{\mathrm{wg}}}\right)^{1/2}\!\!\frac{1}{\sqrt{{\cal V}}}%
\int\!\!d^3\!r\,\mathbf{D}^{(s)\ast}(\mathbf{r})\!\cdot\!\mathbf{e}_{\mathbf{k}\alpha}\,\mathrm{e}^{i\mathbf{k}\cdot\mathbf{r}}%
\nonumber\\%
d_s^\dagger\!&=&\!\sum_{\mathbf{k},\alpha=1,2}\!\!a_{\mathbf{k}\alpha}^{\dagger}\left(\frac{\omega_k^{\mathrm{free}}}{\omega_s^{\mathrm{wg}}}\right)^{1/2}\!\!\frac{1}{\sqrt{{\cal V}}}%
\int\!\!d^3\!r\,\mathrm{e}^{-i\mathbf{k}\cdot\mathbf{r}}\,\mathbf{e}_{\mathbf{k}\alpha}^{\ast}\!\cdot\!\mathbf{E}^{(s)}(\mathbf{r})%
\nonumber\\%
\label{2.9}%
\end{eqnarray}
The expansion (\ref{2.8}) is identical to the basic definition (\ref{2.6}) due to orthogonality and completeness relations for the complete set of the guided and external modes. In expression (\ref{2.9}) we additionally labeled the mode frequencies for distinguishing the field's mode in the presence of the waveguide $\omega_s^{\mathrm{wg}}\equiv\omega_s$ and in free space $\omega_k^{\mathrm{free}}\equiv\omega_j=c\,k$.

Because of difference in mode representation in terms of the electric and the displacement fields the operators $b_s$ and $d^\dagger_s$ are not hermitian conjugated counterparts. Nevertheless they are candidates for respectively annihilation and creation operators of a photon in a specific waveguide mode. It may seem that the difference between $\omega_s^{\mathrm{wg}}$ and $\omega_k^{\mathrm{free}}$ prevents their commutation relations to fulfill the standard bosonic operators. But for the waveguide, designed as a dielectric nanofiber with diameter less than wavelength of the guided light, the free space modes with $\omega_k^{\mathrm{free}}\simeq\omega_s^{\mathrm{wg}}$ mainly dominate in the overlapping integral in (\ref{2.9}) such that the respective factor vanishes in these expansions. Then we can safely accept that operators $b_s$ and $d^\dagger_s$ obey the standard bosonic commutation rules $[b_s,d_{s'}^{\dagger}]=\delta_{s,s'}$. To prove this statement it is important to take into account that the mode of displacement field $\mathbf{D}^{(s)}(\mathbf{r})$ has transversal profile in the reciprocal $k$-representation because of $\mathrm{div}\mathbf{D}^{(s)}(\mathbf{r})=0$.

For the considered process in the case of a near resonant scattering the matrix element of the transition operator (\ref{2.3}) can be disclosed by the following expansion
\begin{eqnarray}
\lefteqn{T_{g's',g\,s}(E)=2\pi\hbar\sqrt{\omega_{s'}\omega_s}}
\nonumber\\%
&&\times\sum_{b,a=1}^{N}\;\sum_{n',n}%
\left(\mathbf{d}\!\cdot\!\mathbf{D}^{(s')}(\mathbf{r}_b)\right)_{n'm'_b}^{*}\left(\mathbf{d}\!\cdot\!\mathbf{E}^{(s)}(\mathbf{r}_a)\right)_{nm_a}%
\nonumber\\%
&&\times\langle\ldots m'_{b-1},n',m'_{b+1}\ldots |\tilde{\hat{R}}(E)%
|\ldots m_{a-1},n,m_{a+1}\ldots \rangle%
\nonumber\\%
&&\label{2.10}%
\end{eqnarray}
where  $\omega_s$ and $\omega_{s'}$ are the frequencies of incident and scattered photons respectively. The transition amplitude is intrinsically determined by the matrix element of the resolvent operator of the system Hamiltonian projected onto a collective atomic state with a single optical excitation
\begin{equation}
\tilde{\hat{R}}(E)=\hat{P}\,\hat{R}(E)\,\hat{P}\equiv \hat{P}\frac{1}{E-\hat{H}}\hat{P}%
\label{2.11}%
\end{equation}
The projector $\hat{P}$ is given by
\begin{eqnarray}
\lefteqn{\hspace{-0.8cm}\hat{P}=\sum_{a=1}^{N}\;\sum_{\{m_j\},j\neq a}\;\sum_{n}%
|m_1,\ldots,m_{a-1},n,m_{a+1},\ldots m_N\rangle}%
\nonumber\\%
&&\hspace{-0.5cm}\langle m_1,\ldots,m_{a-1},n,m_{a+1},\ldots m_N|\times|0\rangle\langle 0|_{\mathrm{Field}}%
\label{2.12}%
\end{eqnarray}
and selects in the atomic Hilbert subspace the entire set of the states where any $j$-th of $N-1$ atoms populates a Zeeman sublevel $|m_j\rangle$ in its ground state and one specific $a$-th atom (with $a$ running from $1$ to $N$ and $j\neq a$) populates a Zeeman sublevel $|n\rangle$ of its excited state. The field subspace is projected onto its vacuum state and operator $\tilde{\hat{R}}(E)$ can be further considered as a matrix operator acting only in atomic subspace.

In the representation of the $T$-matrix by the expansion (\ref{2.10}) the selected specific product of matrix elements runs all the possibilities when the incoming photon is absorbed by any $a$-th atom and the outgoing photon is emitted by any $b$-th atom of ensemble, including the possible coincidence $a=b$. The initial atomic state is given by $|g\rangle\equiv|m_1,\ldots,m_N\rangle$ and the final atomic state by $|g'\rangle\equiv|m'_1,\ldots,m'_N\rangle$, where atoms can populate all the accessible internal states. Thus for the system consisting of many atoms with degenerate ground state there is an exponentially rising number of scattering channels. Nevertheless for most problems, such as quantum memories, the elastic scattering channel is mostly important and can be calculated once we find the resolvent operator (\ref{2.11}).

As can be verified the arbitrary parameter ${\cal L}$ vanishes when substituting (\ref{2.10}) into (\ref{2.7}) such that the $S$-matrix becomes a regular and fairly defined physical quantity in our calculation scheme. Its matrix elements give us the quantum transition amplitudes for observing the system in particular final states in the considered quasi-one-dimensional scattering process. We now turn to the determination of the resolvent operator via  Feynman diagram method in the perturbation theory technique.

\subsection{The resolvent operator}\label{Subsection_II.C}

\noindent Below we apply a microscopic calculation of the projected resolvent operator for an atomic system with degenerate ground state. This approach has been earlier developed in \cite{Sheremet12,Sheremet15,Kupriyanov17} for light scattering in free space and we adapt it here to the waveguide configuration. The inverse resolvent matrix can be expressed in the following form
\begin{equation}
\tilde{\hat{R}}^{-1}(E)=E-\hbar\omega_0-\Sigma(E)%
\label{2.13}
\end{equation}
where $\omega_0$ is the resonant atomic frequency and $\Sigma(E)$ is the self energy part. This term is calculated via its relevant expansion by a set of irreducible diagrams. It is expected to have smooth dependence on its energy argument and for near resonant scattering can be reliably approximated by substituting $E=\hbar\omega_0$ with the assumption that the ground state energy $E_g=0$ for degenerate system of the atomic Zeeman sublevels.

The self-energy part can be constructed by keeping the leading contributions in its diagram expansion. For each $a$-th atom from the ensemble there is the following single-particle self-energy term
\begin{eqnarray}
\lefteqn{\scalebox{1.0}{\includegraphics*{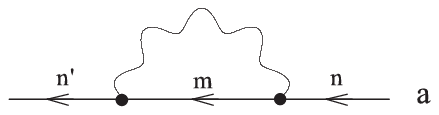}}}
\nonumber\\%
&&\Rightarrow\sum_{m}\int\frac{d\omega}{2\pi} d^{\mu}_{n'm}d^{\nu}_{mn}%
iD^{(E)}_{\mu\nu}(\mathbf{r}_a,\mathbf{r}_a;\omega)%
\nonumber\\%
&&\times\frac{1}{E-\hbar\omega-E_m+i0}%
\equiv\Sigma^{(a)}_{n'n}(E)%
\label{2.14}
\end{eqnarray}
where the internal wavy line expresses the causal-type vacuum Green's function of the chronologically ordered polarization components of the field operators
\begin{equation}
iD^{(E)}_{\mu\nu}(\mathbf{r},\mathbf{r}';\tau)=\left\langle T\hat{E}_{\mu}(\mathbf{r},t)\hat{E}_{\nu}(\mathbf{r}',t')\right\rangle%
\label{2.15}%
\end{equation}
and $D^{(E)}_{\mu\nu}(\mathbf{r},\mathbf{r}';\omega)$ is its Fourier image defined by Eq.(\ref{a.1}) in appendix \ref{Appendix_A}. For the sake of generality we use here covariant notation for vector (dipole) and tensor (Green's function) components and omit the indication of sum over repeated tensor indices. In a particular case of Cartesian frame we will further simplify our notation and show all the tensor components via subscribed indices.

Unlike the similar calculations performed in free space the Green's function (\ref{2.15}) depends here on its spatial arguments separately and is strongly modified in the presence of the waveguide, see appendix \ref{Appendix_A}. As an important consequence, even a single-particle contribution to the self-energy depends on the atom's location and has an anisotropic matrix structure in the basis of its excited states. The respective correction to the atomic energy structure as well as to its decay parameters have anisotropic structure and, in the general case of the degenerate excited state, cannot be simply reduced to the radiative Lamb shift and decay constant. The correct renormalization of the single-particle self-energy part (\ref{2.14}) concerns both the near field self-action as well as the radiative interaction of the atom with the quantized field.

As explained in \cite{Sheremet12} in the first leading orders of the perturbation theory the double particle coupling dominates in the cooperative correction to the self-energy part. The respective contribution is given by the sum of two diagrams where the excitation transfer induced by the dipole-type interaction (\ref{2.5}) have different time ordering. The time ordered transfer of a single optical excitation from an atom $a$ to an atom $b$ is described by the following diagram
\begin{eqnarray}
\lefteqn{\scalebox{1.0}{\includegraphics*{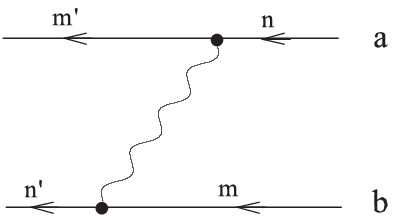}}}
\nonumber\\%
&&\hspace{-0.5cm}\Rightarrow\int\frac{d\omega}{2\pi} d^{\mu}_{n'm}d^{\nu}_{m'n}%
iD^{(E)}_{\mu\nu}(\mathbf{r}_b,\mathbf{r}_a;\omega)%
\nonumber\\%
&&\hspace{-0.5cm}\times\frac{1}{E-\hbar\omega-E_m-E_{m'}+i0}%
\equiv\Sigma^{(ab+)}_{m'n';nm}(E)
\label{2.16}
\end{eqnarray}
and the time anti-ordered transfer by the complementary diagram
\begin{eqnarray}
\lefteqn{\scalebox{1.0}{\includegraphics*{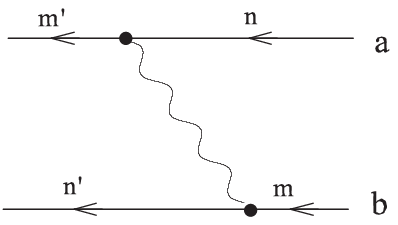}}}
\nonumber\\%
&&\hspace{-0.5cm}\Rightarrow\int\frac{d\omega}{2\pi} d^{\mu}_{n'm}d^{\nu}_{m'n}%
iD^{(E)}_{\mu\nu}(\mathbf{r}_b,\mathbf{r}_a;\omega)%
\nonumber\\%
&&\frac{1}{E+\hbar\omega-E_n-E_{n'}+i0}%
\equiv\Sigma^{(ab-)}_{m'n';nm}(E)
\label{2.17}
\end{eqnarray}
The vector components of the dipole matrix elements $d^{\nu}_{m'n}$ and $d^{\mu}_{n'm}$ are related with the atoms $a$ and $b$ respectively. In the pole approximation $E\approx E_n=\hbar\omega_0$ the delta-function singularities dominate in the sum of spectral integrals (\ref{2.16}) and (\ref{2.17}) and the sum of both terms gives
\begin{eqnarray}
\Sigma^{(ab)}_{m'n';nm}(E)&\approx&\Sigma^{(ab+)}_{m'n';nm}(\hbar\omega_0)+%
\Sigma^{(ab-)}_{m'n';nm}(\hbar\omega_0)%
\nonumber\\%
&=&\frac{1}{\hbar}\,d^{\mu}_{n'm}d^{\nu}_{m'n}\,D^{(E)}_{\mu\nu}(\mathbf{r}_b,\mathbf{r}_a;\omega_0)%
\label{2.18}%
\end{eqnarray}
The derived expression has a clear physical meaning. The real component of the double particle contribution to the self-energy part reproduces both the static interaction between two proximal dipoles and radiative correction to the quasi-energy structure for the distant dipoles. Its imaginary component is responsible for cooperative dynamics of the excitation decay in the entire radiation process. For long distances, when the atomic dipoles are separated by the radiation zone, this latter term describes radiation interference between any pair of two distant atoms, which weakly reduces with the interatomic separation for interaction via external field modes. But for collection of atoms arrayed along the waveguide there is always strong cooperative interaction via evanescent field of the fundamental waveguide mode (see Appendix \ref{Appendix_A}). Thus in the considered quasi-one-dimensional scattering the cooperative effects become extremely important as well as the scattering process becomes strongly dependent on a particular configuration of the atomic array.

\section{Results:Scattering from a nanofiber-trapped atomic array}\label{Section_III}

\begin{figure}[tp]
{$\scalebox{0.6}{\includegraphics*{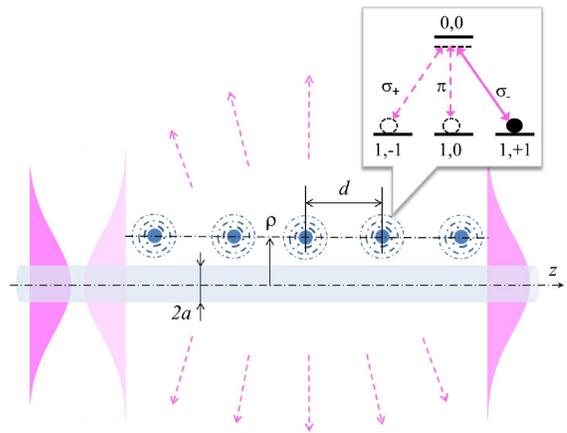}}$}
\caption{(color online). Light scattering from an atomic array trapped near a sub-wavelength dielectric waveguide. The atoms, which are located at a distance $\rho-a$ from the surface, are spin oriented along the waveguide and the incident light is in the left-handed polarized waveguide mode. The scattered light leaves the one-dimensional channel in either forward or backward directions.}
\label{fig1}%
\end{figure}

\noindent In this section we present the results of our numerical simulations for light scattering from an array of atoms trapped near a nanofiber. The geometry is shown in Fig.~\ref{fig1}. The atoms have two energy levels with degenerate Zeeman structure of the ground state. We consider an array of $\Lambda$-configured atoms with the minimal accessible number of quantum states, i.e., with angular momentum $F_0 = 1$ in the ground state and $F =0$ in the excited state. Thus we further associate the quantum indices $m\equiv F_0,M_0$ and $n\equiv F,M=0,0$, where $M_0$ and $M$ are the Zeeman projections of the atomic spin angular momentum of the ground and excited states respectively. Such an energy and angular momentum configuration exists as closed transition in the hyperfine manifold of ${}^{87}$Rb and we use the spectral parameters of rubidium atom in our estimates. We assume the initial collective state of atoms as spin oriented along the waveguide direction such that all the atoms populate only one Zeeman sublevel $F_0 = 1, M_0 = 1$, which is relevant for the realization of quantum interfaces based on atomic systems.

\subsection{The waveguide parameters}
\noindent In our calculations we consider a subwavelength nanofiber with radius $a$ and dielectric constant $\epsilon$. The solution of the homogeneous Maxwell equations can be obtained by factorizing the mode functions of the waveguide in cylindrical coordinates as specified by the first line in  Eq.~(\ref{a.6}) (see \cite{Marcuse} for derivation details). For the fundamental HE$_{11}$ -mode its components can be superposed in the set of three basic functions $E_{\rho}(\rho)$, $E_{\phi}(\rho)$, and $E_{z}(\rho)$, which in turn are compiled by Bessel functions of different type (see Appendix \ref{Appendix_B}).

In cylindrical coordinates the mode components, specified by the first line in Eq.~(\ref{a.6}), are given by
\begin{eqnarray}
E_{\rho}^{(\pm 1 k)}(\rho)&=&E_{\rho}(\rho)%
\nonumber\\%
E_{\phi}^{(\pm 1 k)}(\rho)&=&\pm E_{\phi}(\rho)%
\nonumber\\%
E_{z}^{(\pm 1 k)}(\rho)&=&E_{z}(\rho)%
\label{3.1}%
\end{eqnarray}
and in a Cartesian frame the mode components, specified by the second line in Eq.~(\ref{a.6}), are given by
\begin{eqnarray}
E_{x}^{(\pm 1 k)}(\rho,\phi)&=&\frac{E_{\rho}(\rho)-iE_{\phi}(\rho)}{2\sqrt{2\pi}}+\frac{E_{\rho}(\rho)+iE_{\phi}(\rho)}{2\sqrt{2\pi}}\,\mathrm{e}^{\pm 2i\phi}%
\nonumber\\%
E_{y}^{(\pm 1 k)}(\rho,\phi)&=&\pm\frac{iE_{\rho}(\rho)+E_{\phi}(\rho)}{2\sqrt{2\pi}}\mp\frac{iE_{\rho}(\rho)-E_{\phi}(\rho)}{2\sqrt{2\pi}}\,\mathrm{e}^{\pm 2i\phi}%
\nonumber\\%
E_{z}^{(\pm 1 k)}(\rho,\phi)&=&\frac{E_{z}(\rho)}{\sqrt{2\pi}}\,\mathrm{e}^{\pm i\phi}.%
\label{3.2}%
\end{eqnarray}
For a single-mode waveguide with $a\lesssim k^{-1}$ and with the refraction index $n=\sqrt{\epsilon}\to 1$ the solution approaches to $E_{\rho}(\rho)\to -iE_{\phi}(\rho)$ with vanishing longitudinal component $E_{z}(\rho)\to 0$ such that the fundamental mode becomes independent on the azimuthal angle $\phi$ and can be reliably approximated by a Gaussian fundamental mode propagating in free space. In this paraxial-type approximation the two degenerate modes with $\sigma=\pm 1$  transform to two orthogonal right-handed and left-handed polarized Gaussian modes respectively. However in a more realistic situation, as considered here, with $a\lesssim k^{-1}$ but $n\gtrsim 1$ all the components are competitive and all of them mediate the excitation process in the atomic system.

\begin{figure}[tp]
{$\scalebox{0.8}{\includegraphics*{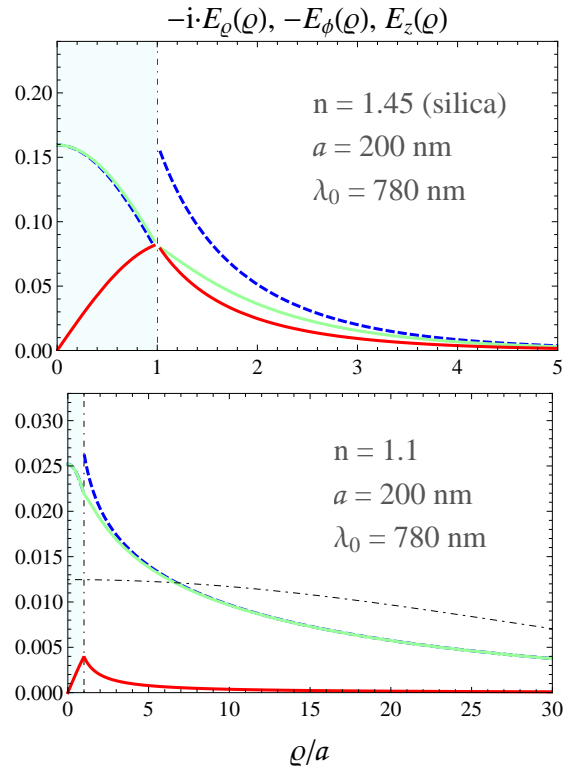}}$}%
\caption{(color online). Functions $-i E_{\rho}(\rho)$ (dashed blue), $-E_{\phi}(\rho)$ (green), and $E_{z}(\rho)$ (red, lower in the graphs) contributing to the waveguide modes (\ref{3.1}) and (\ref{3.2}) for $a~=~200\,$~nm and for vacuum wavelength $\lambda_0=780\, $nm (Rubidium). The vertical shaded area indicates the waveguide. Upper panel displays the mode structure for silica with refraction index $n=1.45$ and lower panel relates to a dielectric medium with $n=1.1$. For this lower panel, we show a Gaussian fit of the fundamental mode (dash-dotted).}
\label{fig2}%
\end{figure}

This important property of a nanofiber is illustrated by the plots shown in Fig.~\ref{fig2}. For a nanofiber with $a=200\,\mathrm{nm}$ and for the mode frequency taken at the rubidium wavelength $\lambda_0=780\,\mathrm{nm}$ we plot three functions $-i E_{\rho}(\rho)$, $-E_{\phi}(\rho)$, and $E_{z}(\rho)$, which can be set as real and which were calculated for two different refraction indices $n=1.45$ (silica) and $n=1.1$ (to follow tendency to the paraxial asymptotic). For the latter case we additionally show the Gaussian fit to the HE$_{11}$-mode to follow how it reproduces the tail asymptote of the evanescent field outside the fiber.

\subsection{Single-atom scattering}
\noindent For a single atom with a non-degenerate upper state its self-energy part (\ref{2.14}) can be expressed as
\begin{equation}
\left.\Sigma(E,\mathbf{r})\right|_{E=\hbar\omega_0}=\hbar\Delta(\mathbf{r})-\frac{i\hbar}{2}\gamma(\mathbf{r})%
\label{3.3}
\end{equation}
where we omitted unnecessary specification by quantum index $n\equiv (F=0,M=0)$ but parameterized the self-energy by the additional argument $\mathbf{r}$ that indicates the atom's position. The real part $\Delta(\mathbf{r})$ is diverging and should be incorporated into the physical energy of the atom. The infinite energy shift is associated with both the dipole self-action and the radiation Lamb-shift (incorrectly described in the dipole gauge), which can be safely renormalized \cite{Kupriyanov17}. Nevertheless, beyond its infinite part $\Delta(\mathbf{r})$ contains a finite and important correction to the energy shift induced by the dipole coupling with the waveguide. Here we avoid the non-trivial part of the calculation of this correction and will further assume that all the atomic dipoles are set in similar conditions such that the unknown quantity $\Delta(\mathbf{r})$ can be associated with an energy shift identical for all dipoles.

According to the expansion of the Green's function in terms of contributions of the guided and external modes (see Eq.~(\ref{a.9}) in Appendix \ref{Appendix_A}), the radiation decay rate $\gamma(\mathbf{r})$ can be similarly expanded as
\begin{equation}
\gamma(\mathbf{r})=-\frac{2}{\hbar^2}\,d_0^2\; \mathrm{Im}\left[D^{(E)}_{\mu\mu}(\mathbf{r},\mathbf{r};\omega_0)\right]=\gamma^{(\mathrm{wg})}(\mathbf{r})+\gamma^{(\mathrm{ext})}(\mathbf{r})
\label{3.4}
\end{equation}
with
\begin{eqnarray}
\gamma^{(\mathrm{wg})}(\mathbf{r})&=&\frac{4\omega_0}{\hbar v_g}\,d_0^2\left[\left|E_{\rho}(\rho)\right|^2+\left|E_{\phi}(\rho)\right|^2+\left|E_{z}(\rho)\right|^2\right]%
\nonumber\\%
\gamma^{(\mathrm{ext})}(\mathbf{r})&=&-\frac{2}{\hbar^2}\,d_0^2\; \mathrm{Im}\left[D^{(\mathrm{ext})}_{\mu\mu}(\mathbf{r},\mathbf{r};\omega_0)\right]%
\label{3.5}%
\end{eqnarray}
where $d_0$ denotes modulus of the dipole matrix element, same for all optical transitions in the considered case, and we conventionally assume sum over the repeated tensor index $\mu=x,y,z$. Because of the axial symmetry the decay rate depends only on the distance $\rho$ of the atom from $z$-axis. The waveguide contribution $\gamma^{(\mathrm{wg})}(\mathbf{r})$ is a result of exact substitution of the respective contribution to the Green's function, given by Eq.~(\ref{a.10}). By approximating $D^{(\mathrm{ext})}$ with (\ref{a.16}) and (\ref{a.17}) we can estimate the contribution of the external modes $\gamma^{(\mathrm{ext})}(\mathbf{r})$ as the following correction to the natural decay constant $\gamma$
\begin{eqnarray}
\lefteqn{\gamma^{(\mathrm{ext})}(\mathbf{r})\sim\gamma}%
\nonumber\\%
&-&\frac{2\omega_0}{\hbar c}\,d_0^2\left[\left|E_{\rho}(\rho)\right|^2+\left|E_{\phi}(\rho)\right|^2+2\mathrm{Re}\left[iE_{\rho}(\rho)E_{\phi}^*(\rho)\right]\right]%
\nonumber\\%
\label{3.6}%
\end{eqnarray}
As commented in Appendix \ref{Appendix_A}, the second term eliminates spontaneous emission into those vacuum modes, which in the paraxial approach coincide with the waveguide modes.

\begin{figure}[tp]
{$\scalebox{0.8}{\includegraphics*{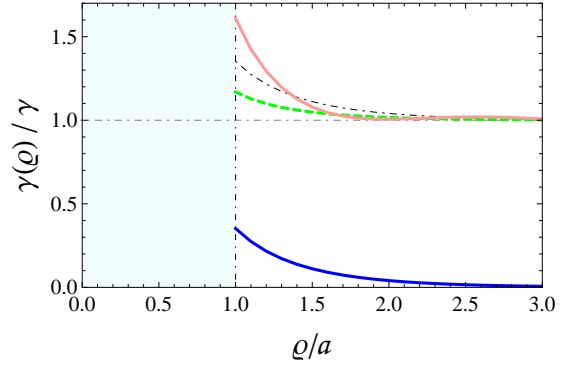}}$}%
\caption{(color online). Spontaneous decay rate $\gamma(\rho)$ of a Rubidium atom located near the waveguide as a function of its distance $\rho-a$ to the surface: (blue, lower curve) contribution of the fundamental mode; (black dashed-dotted) sum of contributions of the fundamental mode and natural decay $\gamma$; green dashed) calculations based on Eqs.(\ref{3.4}) and (\ref{3.5}); (red) exact result. The vertical shaded area indicates the waveguide. The waveguide parameters are the same as in the upper plot of Fig.~\ref{fig2}.}
\label{fig3}%
\end{figure}

In Fig.~\ref{fig3} we show the results of our numerical simulations for the rate of spontaneous decay $\gamma(\rho)$ as a function of the radial position of the atom from the surface. The calculations, based on Eqs.(\ref{3.5}) and (\ref{3.6}), are performed for silica, with mode functions shown in the upper plot of Fig.~\ref{fig2}, and compared with the exact result calculated via Fermi's golden rule where the complete set of the external modes is kept \cite{Hakuta2005}, \cite{Balykin2004_field_distribution}. As can be seen from the plotted graphs the highest deviation is observed when the atom is located near the fiber surface. That indicates a significant contribution from the process of recurrent scattering to the Green's function, which can be recovered via iterative solution of the scattering equation (\ref{a.12}) in Appendix \ref{Appendix_A}. Nevertheless at the distance comparable with the waveguide radius $a$ the exact result is relevantly reproduced by estimate (\ref{3.6}). It is noteworthy to point out that at the intermediate distances, where $\rho-a\sim a$, the simple sum of the natural $\gamma$ with the waveguide contribution overestimates $\gamma(\rho)$ and should be corrected as suggested by Eq.~(\ref{3.6}).

\begin{figure}[tp]
{$\scalebox{0.75}{\includegraphics*{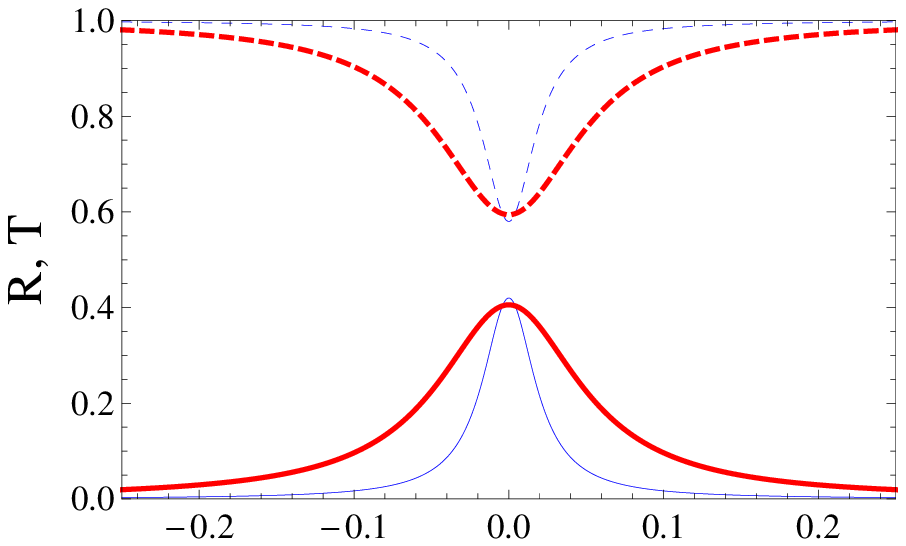}}$}\\%
{$\scalebox{0.75}{\includegraphics*{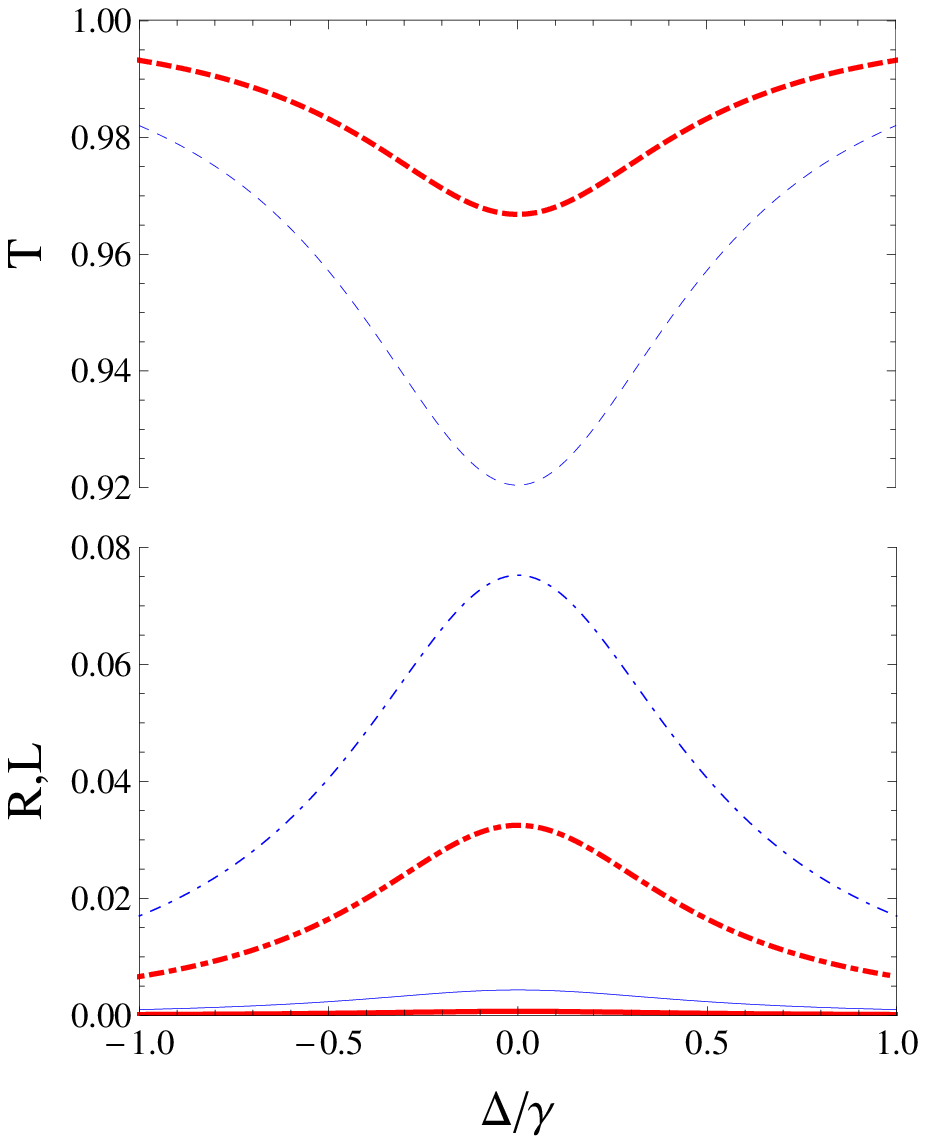}}$}
\caption{(color online). The transmission $\mathrm{T}$ (dashed lines), reflection $\mathrm{R}$ (solid lines) and losses $\mathrm{L}=1-\mathrm{R}-\mathrm{T}$ (dash-dotted lines) calculated as a function of frequency detuning $\Delta=\omega-\omega_0$ for light scattering from a single Rubidium atom trapped at distances $\rho-a= 0.5\, a$ (blue thin lines) and $\rho-a= a$ (red thick lines). The mode and atomic decay parameters are the same as in Figs.~\ref{fig2} and \ref{fig3}. The upper pannel provides the spectra for the lossless scattering with light emission into the waveguide mode only while the lower plot corresponds to the complete scattering process.}
\label{fig4}%
\end{figure}

The obtained parameters contribute to the resolvent operator and $S$-matrix as explained in section \ref{Section_II}.  The light scattering in a quasi-one-dimensional geometry can be described by coefficients of transmission $\mathrm{T}$, reflection $\mathrm{R}$ and losses $\mathrm{L}$, which are respectively given by
\begin{eqnarray}
\mathrm{T}&=&\mathrm{T}(\omega)=\sum_{i',k'>0}\left|S_{i'i}\right|^2%
\nonumber\\%
\mathrm{R}&=&\mathrm{R}(\omega)=\sum_{i',k'<0}\left|S_{i'i}\right|^2%
\nonumber\\%
\mathrm{L}&=&\mathrm{L}(\omega)=1-\mathrm{R}(\omega)-\mathrm{T}(\omega)
\label{3.7}%
\end{eqnarray}
where $i=\{\sigma=-1,k;M_0=+1\}$ and sum over $i'=\{\sigma',k';M'_0\}$ includes both polarization channels $\sigma'=\pm 1$ and three atomic transitions with $M'_0=0,\pm 1$. The elastic forward and backward scattering channels are distinguished by the signs of the longitudinal wave numbers $k'=+k$ and $k'=-k$ respectively, see Eq.~(\ref{a.6}). The wave number of an incident photon is parameterized by its frequency accordingly to the dispersion law of the waveguide $k=k(\omega)$. In Fig.~\ref{fig4} we plot the single-atom transmission and reflection coefficients as function of frequency detuning from atomic resonance $\Delta=\omega-\omega_0$ for two different positions of the atom near a nanofiber: $\rho-a=0.5\, a$ and $\rho-a= a$. The waveguide parameters are chosen the same as in the upper plot of Fig.~\ref{fig2}. The upper plot of Fig.~\ref{fig4} shows the result of our calculations in assumption that the atom would emit the light into the waveguide mode only. For such an ideal lossless configuration the balance $\mathrm{R}(\omega)+\mathrm{T}(\omega)=1$ evidently fulfills in all the spectral domain. For complete scattering process the graphs of the lower plot also indicate a few-percents interaction of the atom with the evanescent field. The small but not negligible fraction of light is mostly scattered into external modes and the scattering process is as strong as the atom is closer to the fiber surface. In the next section we study the scattering provided by a full chain of atoms trapped along the waveguide.

\subsection{Light scattering from an atomic array}

\noindent We consider the scattering process from a collection of atoms trapped along the waveguide in the geometry shown in Fig.~\ref{fig1}. In the case of several atoms the scattering problem is described by the same set of parameters given by Eq.~(\ref{3.7}) with updated definitions of initial and final states. Initially our system is prepared in a spin oriented collective state such that $i=\{\sigma=-1,k;M_0=+1 (\mathrm{all\ atoms})\}$ but the final state can be any of $i'=\{\sigma',k';{\{M^{(a)\prime}_0\}}_{a=1}^N\}$ where each of the $N$ atoms can be redistributed onto arbitrary Zeeman sublevel with $M^{(a)\prime}_0=0,\pm 1$. The dimension of the Hilbert space for the resolvent operator as well as the number of scattering channels exponentially rises up with the number of atoms. However, as confirmed by our numerical simulations, for an array consisting of relatively small number of atoms the Rayleigh channel makes dominant contribution to the scattering process. Contribution of the Raman process can be reliably estimated by keeping in the output channel only a single spin flip equally shared among all atoms of the ensemble.

\subsubsection{Ordered atomic array}

\begin{figure}[tp]
{$\scalebox{0.8}{\includegraphics*{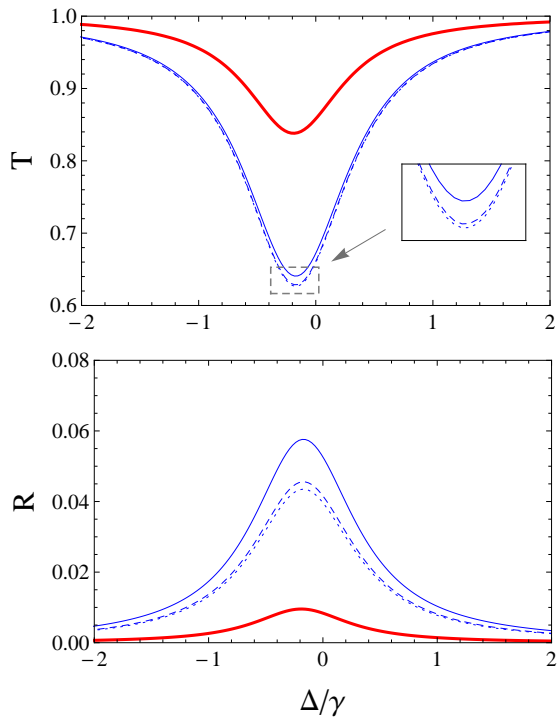}}$}
\caption{(color online). The transmission $\mathrm{T}$ (upper plot) and reflection $\mathrm{R}$ (lower plot) calculated for an array of five ordered Rubidium atoms trapped at distances $\rho-a= 0.5\, a$ (blue thin lines) and $\rho-a= a$ (red thick lines). The atoms are separated by a distance $d=\pi/k=\lambda^{\mathrm{wg}}/2$. For $\rho-a= 0.5\, a$ we additionally indicate the partial contributions of Rayleigh scattering channels with preserving mode polarization (dotted) and with keeping both the polarization components of the outgoing light (dashed). The solid curves show the total contribution including Raman scattering channels. Other parameters are the same as in Fig.~\ref{fig4}.}
\label{fig5}%
\end{figure}

\noindent First we simulate an ordered array of atoms. Fig.~\ref{fig5} provides the parameters of the scattering process calculated for a system of five atoms with a fixed longitudinal separation $d=\pi/k=\lambda^{\mathrm{wg}}/2$ (half of the mode wavelength, see Eq.~(\ref{a.6})), as it is shown in Fig.~\ref{fig1}. Similarly to the single-atom case the array is considered as located at two distances $\rho-a= 0.5\, a;\ a$ from the fiber surface. For the shortest distance we additionally indicate the partial contributions of Rayleigh scattering channel, which leaves atoms at the initial Zeeman state, but either preserves mode polarization or can transfer the outgoing light into orthogonal polarization mode. As it can be seen the Rayleigh contribution dominates in the entire scattering process. As pointed out above the Raman scattering mainly results from a single spin flip, shared among all atoms of the ensemble, such that most of the atoms preserve their initial population of $F_0 = 1, M_0 = 1$ state. This can be explained by approximate azimuthal symmetry of the complete system (photon and atoms) in respect to collinear geometry of the forward scattering. The symmetry provides conservation for the total angular momentum such that the spin angular momentum transfer physically implies the mechanism of spin exchange between the photon and atomic spin subsystem in the Raman process, which just results in a single spin flip. It is also noteworthy to point out that for the Rayleigh channel the scattering into the orthogonal polarization mode, i. e. into $\sigma'=+1$ for forward or $\sigma'=-1$ for backward directions, is possible but quite small for both the transmission and reflection. Such scattering channel would be completely forbidden in the paraxial approach. For the considered small collection of atoms the reflection is still weak and the losses associated with incoherent scattering can be mainly estimated by deviation of the transmission coefficient from the level of ideal transparency.

The transmission $\mathrm{T}(\omega)$ and the reflection $\mathrm{R}(\omega)$ spectra from the atomic array now demonstrate a clear signature of cooperativity in the scattering process. By comparing the spectral dependencies of Fig.~\ref{fig4} and Fig.~\ref{fig5} we can see that the incoherent losses are enhanced by the factor of $N=5$ in accordance with the natural tendency to the Beer-Lambert law. However the reflection exhibits much stronger enhancement roughly scaled by a factor of $N^2=25$ and justifies the effect of coherent Bragg-type reflection. Such scaling is valid only for a small number of scatterers and far from the saturation limit. In the case of mesoscopic system the enhancement from an atomic array consisting of many atoms, showing strong reflection and the effect of one-dimensional atomic mirror, has been recently observed \cite{Corzo2016,Polzik16}. Let us emphasize that in our approach this observation rigorously results from the ab-initio calculation of the scattering process, which is based on the resolvent operator and the $S$-matrix formalism. Interpretation of the Bragg reflection via a semiclassical approach and the transfer matrix formalism can be found in the above references.

All the spectral dependencies are red-shifted from the atomic resonance and from the single-atom spectra presented in Fig.~\ref{fig4}. This is a consequence of the short range static interaction of atomic dipoles and is a precursor of the well-known Lorentz-Lorenz effect existing in infinite, homogeneous and dense dielectric media. This type of interaction is naturally incorporated into our calculation scheme as explained in the previous section and detailed in Appendix \ref{Appendix_A}. Due to the translational symmetry in the lattice-type and ordered atomic configuration, all the presented spectra have Lorentzian-shaped monotonic profiles. Such a smoothed spectral behavior would be dramatically changed once we introduce disorder in the atomic distribution, such as delocalization of the trapping potential wells due to finite temperature.

\subsubsection{Disordered atomic array}

\begin{figure}[tp]
{$\scalebox{0.8}{\includegraphics*{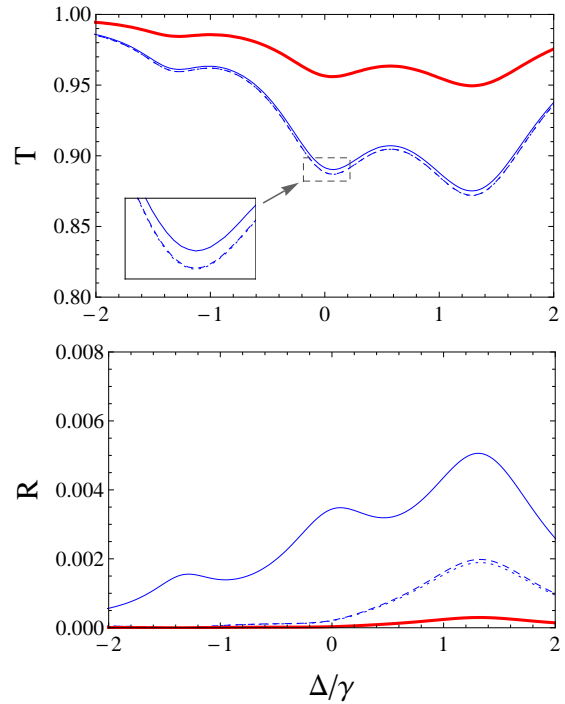}}$}
\caption{(color online). Reflection and transmission spectra as in Fig.~\ref{fig5} but here for a particular random configuration of the atoms in a disordered array.}
\label{fig6}%
\end{figure}

\noindent Figure \ref{fig6} presents the modified spectra for the case of disordered configuration. The atoms have the same average separation $d\sim\pi/k=\lambda^{\mathrm{wg}}/2$, but are randomly and uniformly distributed along the chain. The spectral dependencies, plotted for a particular configuration, become quite sensitive to its variation because of internal correlations in the entire system. The cooperative effects are revealed in a different way and the scattering process is generally weaker. This is clearly seen in the backward scattering channel whose outcome is an order of magnitude smaller than in the case of ordered array, and is negligible when compare with the incoherent losses. Another important feature of the scattering process is the complicated structure of the transmission spectra where several local minima and maxima appear. This is a signature of a microcavity structure created by a disordered but cooperatively organized system composed of atomic scatterers in a one-dimensional configuration \cite{SKKH2009}. In the case of disorder, the cooperativity tends to an Anderson-type localization mechanism that suppresses the scattering process in the one-dimensional geometry.

It is important to point out that such a configuration-sensitive interference for an impinging photon scattered from either ordered or disordered atomic chains has here a rigorous microscopic description based on the structure of the resolvent operator of the system Hamiltonian (\ref{2.11}). Indeed, a single-photon state of the electromagnetic field is an intrinsically quantum microscopic object, which has no phase, is parameterized by a negative-valued Wigner function in its phase space, and cannot be fairly introduced in classical optics. Although the empirical description of the process in terms of the classical wave scattering (i.e. in terms of scattering of a weak coherent light) with classical interpretation of interference paths may give a realistic estimate of the output transmission and reflection spectra, it should be provided of the phenomenologically defined elementary scattering parameters namely by a single-atom decay constant and scattering amplitude. In this sense the performed calculations emphasize that the resolvent operator (\ref{2.11}), transformed to its diagonal form, generates the set of unstable Dicke-type excited quantum entangled states, which have strong internal and configuration-sensitive correlations, and which respond to the driving photon as one complex quantum system. The parameters of the system are rigorously defined by its self-energy part (effective Hamiltonian), introduced in Section \ref{Subsection_II.C}. The microscopically calculated resolvent operator correctly reproduces the complicated dynamics of the scattering process and its sensitivity to variation of the external conditions.

\section{Conclusion}\label{Section_IV}
\noindent In this paper we have developed the principles of quantum scattering theory toward the microscopic description of cooperative light scattering from an array of atoms trapped near a sub-wavelength dielectric waveguide. The developed approach is entirely applicable for a single-photon scattering in a quasi-one-dimensional geometry from atoms having multilevel and degenerate energy structure. The basic mathematical attributes of the scattering process, namely the $S$-matrix and resolvent operator, are linked to the transmission and reflection coefficients characterizing the propagation of the guided photon through the atomic chain. The crucial elements of the performed calculation scheme are the self-energy part and the electric field Green's function and both are strongly affected by the presence of the waveguide. In the case of a nanofiber that supports a single fundamental mode, these functions can be analytically constructed within certain approximations as we have described.

In the context of a quantum interface and coherent control of the signal light, atomic systems with degenerate ground state are specifically interesting. For such systems, the dimension of the Hilbert subspace of the quantum scattering equations rises exponentially with the number of atoms. Therefore in the present study we have restricted our numerical simulations to a configuration of five $\Lambda$-configured atoms with the minimal accessible number of quantum states, i.e. with angular momentum $F_0 = 1$ in the ground state and $F =0$ in the excited state. For this case we have provided an important illustration that includes not only Rayleigh but also Raman channels in microscopic description of the cooperative scattering process in the correct vector model by keeping the complete angular momentum structure of the guided light.

The atomic scatterers have been considered as distributed in either ordered or disordered arrays, and the theory predicts that the parameters of the scattering process strongly depend on the distribution type. Our numerical simulations show that in the case of an atomic chain structured as one-dimensional lattice with a period of half-wavelength of the guided mode there is a significant enhancement of the backward scattering and light reflection. Importantly, this clear manifestation of the Bragg diffraction in the scattering process is obtained here as result of ab-initio description via the calculation of the resolvent operator and can be associated with specific periodic entangled structure of its eigenstates created by an optical excitation. In the alternative situation where the atoms are distributed randomly with random phase-matching conditions for an optical excitation shared among the atoms, we have observed strong dependence of the forward scattering and light transmission on a particular atomic configuration. The transmission spectrum has a non-monotonic profile with several configuration-sensitive maxima and minima, which indicates a certain precursor of the Anderson-type localization mechanism in the conditions of quasi-one-dimensional scattering geometry.

The developed approach can be further generalized and applicable for description of a various light-matter interface protocols developed in ensembles consisting of a macroscopic number of atoms. Potentially this can be done because of convenient approximate form of the electric field Green's function, which we have found in the paper. The main difficulty is in an exponentially uprising dimension of the Hilbert space in the case of a many particle problem. But this difficulty could be overcome by involving mainly the proximate dipoles in the calculation of the self-energy part similarly to how it was demonstrated for atomic systems in free space, see \cite{Sheremet15,Kupriyanov17}.

\section*{Acknowledgements}

\noindent This work was supported by the RFBR grant 15-02-01060, the Emergence program from Ville de Paris, the IFRAF DIM NanoK from R\'egion Ile-de-France and
the PERSU program from Sorbonne Universit\'{e}s. N.V.C. and A.S.Sh. are supported by the EU (Marie Curie Fellowship). The work was carried out with financial support from the Ministry of Education and Science of the Russian Federation in the framework of increase Competitiveness Program of NUST "MISIS", implemented by a governmental decree dated 16th of March 2013, No 211.

\appendix

\section{The electric field Green's function near a sub-wavelength dielectric waveguide}\label{Appendix_A}

\noindent In this appendix we consider how the microscopic Green's function of the electric field is modified in the presence of a sub-wavelength dielectric waveguide. The waveguide, designed as a tiny nanoscale optical fiber ("nanofiber") produced from transparent dielectric medium (silica), can support propagation of only one (confined in the transverse direction) fundamental mode, conventionally named as $\mathrm{HE}_{11}$ -mode. We aim to find the relevant correction to the microscopic Green's function outside the nanofiber, which is associated with the modified structure of the field's modes.

\subsection{Basic links with macroscopic theory}

\noindent As proven in statistical physics, see \cite{LfPtIX}, the causal-type electric field Green's function considered in a spatial region nearby a macroscopic object can be expressed by the retarded-type fundamental solution of the macroscopic Maxwell equation
\begin{eqnarray}
D^{(E)}_{\mu\nu}(\mathbf{r},\mathbf{r}';\omega)\!\!&=&\!\!-i\!\int^{\infty}_{-\infty}\!\!d\tau\,%
\mathrm{e}^{i\omega\tau}\!\left.\langle T E_{\mu}(\mathbf{r},t)\,E_{\nu}(\mathbf{r}',t')\rangle\right|_{\tau=t-t'}%
\nonumber\\%
\!&=&\!\frac{\omega^2}{c^2}D^{(R)}_{\mu\nu}(\mathbf{r},\mathbf{r}';|\omega|)%
\label{a.1}
\end{eqnarray}
Here the integrand under the Fourier transform, introduced by Eq.(\ref{2.15}) as an element of the diagram expansion of the self-energy part, performs the expectation value of the time-ordered product between the exact field operators in the Heisenberg representation, "dressed" by the interaction with the object. The second line identifies this quantity as the retarded-type fundamental solution of the macroscopic Maxwell equation (photon propagator in a medium \cite{LfPtIX})
\begin{eqnarray}
\lefteqn{\triangle D^{(R)}_{\mu\nu}(\mathbf{r},\mathbf{r}';\omega) -\frac{\partial^2}{\partial x_{\mu}\partial x_{\alpha}}D^{(R)}_{\alpha\nu}(\mathbf{r},\mathbf{r}';\omega)}
\nonumber\\%
&+&\frac{\omega^2}{c^2}\left[1+4\pi\chi(\mathbf{r})\right]D^{(R)}_{\mu\nu}(\mathbf{r},\mathbf{r}';\omega)=4\pi\hbar\,\delta_{\mu\nu}\delta(\mathbf{r}-\mathbf{r}')%
\nonumber\\%
\label{a.2}
\end{eqnarray}
where $\chi(\mathbf{r})$ is the spatially dependent dielectric susceptibility of the medium. In the vacuum case, when $\chi(\mathbf{r})\to 0$ expressions (\ref{a.1}) and (\ref{a.2}) reproduce the direct relation between electric field Green's function and photon propagator in free space. But in the general case, they allow us to obtain the correction to the Green's function due to a nearby macroscopic object such as a dielectric waveguide. By neglecting dispersion as we can assume if $\omega$ varied in a narrow spectral domain near the reference atomic resonance frequency $\omega_0$, equation (\ref{a.2}) can be re-written directly for the positive frequency component of the causal-type Green's function $D^{(E)}$ just by adding the factor $\omega^2/c^2$ in its right-hand side.

In this case, and for a transparent medium the standard quantization scheme in free space can be straightforwardly generalized with tracking boundary conditions on the object surface. Expansions (\ref{2.8}) and (\ref{2.9}) in the main text construct the field components as Schr\"{o}dinger operators re-expanded in the complete set of modified basic operators for creation and annihilation of a photon either in the modes confined with the waveguide or in the delocalized external modes. The expectation value of the time ordered product for such modified field operators in the Heisenberg representation automatically reproduces the basic relations (\ref{a.1}) and (\ref{a.2}).

The obtained differential equation for the causal-type Green's function can be further transformed to the following integral form
\begin{eqnarray}
\lefteqn{D^{(E)}_{\mu\nu}(\mathbf{r},\mathbf{r}';\omega) = D^{(0)}_{\mu\nu}(\mathbf{r}-\mathbf{r}';\omega)}
\nonumber\\%
&-&\frac{\Lambda}{\hbar}\int\!d^{3}r''\,D^{(0)}_{\mu\alpha}(\mathbf{r}-\mathbf{r}'';\omega)\chi(\mathbf{r}'')D^{(E)}_{\alpha\nu}(\mathbf{r}'',\mathbf{r}';\omega)%
\nonumber\\%
\label{a.3}
\end{eqnarray}
where the Green's function of the freely propagating field (not modified by the waveguide) is given by
\begin{eqnarray}
\lefteqn{\hspace{-1.5cm}D^{(0)}_{\mu\nu}(\mathbf{R};\omega)\!\!=\!\!-i\!\int^{\infty}_{-\infty}\!\!d\tau\,%
\mathrm{e}^{i\omega\tau}\!\left.{\langle T E_{\mu}\!(\mathbf{r},t)\,E_{\nu}\!(\mathbf{r}',t')\rangle}^{(0)}\right|_{\scriptsize\begin{array}{c}\tau\!=\!t\!-\!t'\\ \mathbf{R}\!=\!\mathbf{r}\!-\!\mathbf{r}'\end{array}}}%
\nonumber\\%
&=&-\hbar\frac{|\omega|^3}{c^3}\left\{i\frac{2}{3}h^{(1)}_0\left(\frac{|\omega|}{c}R\right)\delta_{\mu\nu}\right.%
\nonumber\\%
&+&\left.\left[\frac{X_{\mu}X_{\nu}}{R^2}-\frac{1}{3}\delta_{\mu\nu}\right]%
ih^{(1)}_2\left(\frac{|\omega|}{c}R\right)\right\}%
\label{a.4}
\end{eqnarray}
where the upscribed zero index emphasizes that the averaging is done in the basis of the plane-wave modes. Here $h^{(1)}_L(\ldots)$ with $L=0,2$ are the spherical Hankel functions of the first kind. In the integral equation (\ref{a.3}) we have added an auxiliary parameter $\Lambda\to 1$ for resolving a possible conflict with the Fredholm alternative.

Indeed $\Lambda= 1$ is an eigenvalue of the integral operator in Eq.~(\ref{a.3}) and the respective eigenfunction (integrable in the transverse plane) is given by the solution of the homogeneous wave equation
\begin{eqnarray}
\triangle E^{(s)}_{\mu}(\mathbf{r})\!-\!\frac{\partial^2}{\partial x_{\mu}\partial x_{\alpha}}E^{(s)}_{\alpha}(\mathbf{r})\!%
+\!\frac{\omega^2}{c^2}\left[1+4\pi\chi(\mathbf{r})\right]E^{(s)}_{\mu}(\mathbf{r})\!\!&=&\!\!0%
\nonumber\\%
\label{a.5}
\end{eqnarray}
as can be straightforwardly verified by applying the wave operator to Eq.~(\ref{a.3}). This is a signature of a resonance contribution of the waveguide modes, determined as solutions (localized in the transverse plane) of Eq.~(\ref{a.5}), to the Green's function.

If $\chi(\mathbf{r})\sim \mathrm{const}$ inside the fiber then due to azimuthal and translational symmetry the solution can be factorized in the following product
\begin{eqnarray}
E^{(s)}_{q}(\mathbf{r})&=&E^{(\sigma k)}_{q}(\rho)\,\frac{1}{\sqrt{2\pi {\cal L}}}\mathrm{e}^{i\sigma\phi}\,\mathrm{e}^{ik z}\ \ \ (\mathrm{cylindric})%
\nonumber\\%
E^{(s)}_{\mu}(\mathbf{r})&=&E^{(\sigma k)}_{\mu}(\rho,\phi)\,\frac{1}{\sqrt{{\cal L}}}\mathrm{e}^{ik z}\ \ \ \ \ (\mathrm{Cartesian})%
\label{a.6}
\end{eqnarray}
where the longitudinal wave number $k$ and azimuthal quantum number $\sigma=\pm 1$ are the mode parameters incorporated in the entire mode index $s=\sigma,k$. The solution is obtained in cylindrical coordinates $\mathbf{r}\to \rho,\phi,z$ but with the vector projection defined in respect to either cylindrical basis with $q=\rho,\phi,z$ (first line) or to Cartesian frame with $\mu=x,y,z$ (second line). For a sake of convenience, see normalization condition(\ref{a.8}) below, we pose a longitudinal wave inside a certain quantization segment of length ${\cal L}$ with periodic boundary conditions and consider $k=2\pi/{\cal L}\times(\mathrm{any\;integer})$ as a quasi discrete variable. The mode frequency and longitudinal wave number are connected via the dispersion relation $\omega=\omega_s\equiv\omega_{k}$, which depends on the waveguide parameters, such that each particular frequency $\omega$ determines only one specific wave number $k=k(\omega)$.

Next observation is that there is a continuum family of eigenfunctions of integral operator contributing to equation (\ref{a.3}), which has formed (\ref{a.6}) and correspond to a family of eigenvalues $\Lambda=\Lambda_{k}\sim 1$ varied with $k$. This can be justified by the fact that with slightly varying $\Lambda$ we can change the dielectric susceptibility of the waveguide $\chi(\mathbf{r})\to \Lambda\chi(\mathbf{r})$ and can always fit it to the value, that provides provide a homogeneous solution of such modified Eq.(\ref{a.5}) in form (\ref{a.6}) for arbitrary $k$.

For the interaction with atoms, the evanescent field frequency $\omega$ is expected to be quite close to the atomic frequency $\omega_0$. The possible variations of $k$ from $k(\omega)$ can be scaled by variation of frequency detuning as $\Delta\omega/c$, where $\Delta\omega\ll\omega,\omega_0$ by many orders of magnitude.  In this case, the eigenfunctions of the integral operator for any acceptable $k$ have eigenvalue $\Lambda_k$ very close to "one" and, being considered on a macroscopic scale of a finite waveguide, are practically indistinguishable from the waveguide modes (\ref{a.6}) for the same $k$. Excitation of these quasi-resonant waveguide modes makes considerable correction to the electric field Green's function near the nanofiber.

\subsection{Contribution of the waveguide modes}

\noindent Consider equation (\ref{a.2}) for the causal-type Green's function and in the case when both the spatial arguments $\mathbf{r}$ and $\mathbf{r}'$ are located outside the fiber
\begin{eqnarray}
\lefteqn{\triangle D^{(E)}_{\mu\nu}(\mathbf{r},\mathbf{r}';\omega) -\frac{\partial^2}{\partial x_{\mu}\partial x_{\alpha}}D^{(E)}_{\alpha\nu}(\mathbf{r},\mathbf{r}';\omega)}
\nonumber\\%
&+&\frac{\omega^2}{c^2}D^{(E)}_{\mu\nu}(\mathbf{r},\mathbf{r}';\omega)=4\pi\hbar\,\frac{\omega^2}{c^2}\,\delta_{\mu\nu}\,\delta(\mathbf{r}-\mathbf{r}')%
\label{a.7}
\end{eqnarray}
Let us specify the waveguide modes, defined by Eqs.(\ref{a.5}) and (\ref{a.6}), by the following normalization condition
\begin{eqnarray}
\lefteqn{\int\! d^3r\,\epsilon(\mathbf{r})\, \mathbf{E}^{(s')\ast}(\mathbf{r})\cdot\mathbf{E}^{(s)}(\mathbf{r})}
\nonumber\\%
&&\equiv\int\! d^3r\,\mathbf{D}^{(s')\ast}(\mathbf{r})\cdot\mathbf{E}^{(s)}(\mathbf{r})=\delta_{s's}%
\label{a.8}
\end{eqnarray}
where $\epsilon(\mathbf{r})=1+4\pi\chi(\mathbf{r})$ is the dielectric permittivity and $\mathbf{D}^{(s')}(\mathbf{r})=\epsilon(\mathbf{r})\, \mathbf{E}^{(s')}(\mathbf{r})$ is the $s'$-mode of displacement field. As explained above, the integral over the $z$-variable is bounded by the quantization segment ${\cal L}\to\infty$ and implies periodic boundary conditions and quasi-discrete wave number $k$.

The formal solution of equation (\ref{a.7}) can be constructed via an expansion of the Green's function in the series of the complete basis set for all the cylindrical modes i.e. given by the waveguide modes (\ref{a.6}) and by infinite set of the external delocalized modes. Thus the Green's function is given by the sum of two contributions
\begin{equation}
D^{(E)}_{\mu\nu}(\mathbf{r},\mathbf{r}';\omega)=D^{(\mathrm{wg})}_{\mu\nu}(\mathbf{r},\mathbf{r}';\omega)+D^{(\mathrm{ext})}_{\mu\nu}(\mathbf{r},\mathbf{r}';\omega)
\label{a.9}
\end{equation}
The first term specifies contribution of the waveguide modes and outside the fiber, where the modes of displacement and electric fields coincide, is given by
\begin{equation}
D^{(\mathrm{wg})}_{\mu\nu}(\mathbf{r},\mathbf{r}';\omega)=\sum_{s}\frac{4\pi\hbar\omega^2}{\omega^2-\omega_{s}^{2}+i0}E^{(s)}_{\mu}(\mathbf{r})\,E^{(s)\ast}_{\nu}(\mathbf{r}')%
\label{a.10}
\end{equation}
Contribution of the external modes, denoted by the second term in (\ref{a.9}), can be similarly obtained but, by keeping all the modes, it makes it a quite cumbersome and hardly applicable result for evaluating the cooperative resolvent operator in a many particle problem. Nevertheless, as we show below, in the case of a single mode nanofiber we can modify the integral equation (\ref{a.3}) to another form, which suggests convenient approximation for $D^{(\mathrm{ext})}$.

\subsection{Contribution of the external modes}

\noindent In expansions (\ref{a.9}) and (\ref{a.10}) the spatial argument $\mathbf{r}$ can be taken even inside the fiber by accepting the Green's function as a fundamental solution of the wave equation in its general form (\ref{a.2}). Then, due to the orthogonality of the basis functions associated with the waveguide, we can select the contribution of external modes via the following identity
\begin{eqnarray}
\lefteqn{\hspace{-0.7cm}D^{(\mathrm{ext})}_{\mu\nu}(\mathbf{r},\mathbf{r}';\omega)=D^{(E)}_{\mu\nu}(\mathbf{r},\mathbf{r}';\omega)}%
\nonumber\\%
&&\hspace{-1cm}-\sum_{s}E^{(s)}_{\mu}(\mathbf{r})\!\int\!\!d^3r''\,D^{(s)\ast}_{\alpha}(\mathbf{r}'')D^{(E)}_{\alpha\nu}(\mathbf{r}'',\mathbf{r}';\omega)%
\label{a.11}
\end{eqnarray}
The second term actually subtracts the contribution (\ref{a.10}), performed here in the projected form. If the waveguide is excited on frequency $\omega$ it would be effectively responding on those modes whose frequencies $\omega_s\equiv\omega_k$ are close to $\omega$. Having also in mind that the frequency $\omega$ is quasi-resonant to the reference atomic frequency $\omega_0$  we can expect that only a limited number of near resonant modes with $\omega_k\sim\omega\sim\omega_0$ would make meaningful contribution in (\ref{a.11}). As was pointed above, these modes, considered on a macroscopic scale, of the finite waveguide approximately coincide with the eigenfunctions of the integral operator in equation (\ref{a.3}) with eigenvalue $\Lambda_k\sim 1$.

By substituting expansion (\ref{a.11}) into the integral equation (\ref{a.3}) and by taking $\Lambda=1$ it can be transformed to the following form
\begin{eqnarray}
\lefteqn{D^{(\mathrm{ext})}_{\mu\nu}(\mathbf{r},\mathbf{r}';\omega)\approx D^{(0)}_{\mu\nu}(\mathbf{r}-\mathbf{r}';\omega)}%
\nonumber\\%
&+&\frac{1}{\hbar}\int\!d^{3}r''\,K_{\mu\alpha}(\mathbf{r},\mathbf{r}'';\omega)D^{(\mathrm{ext})}_{\alpha\nu}(\mathbf{r}'',\mathbf{r}';\omega)%
\label{a.12}
\end{eqnarray}
with the following modified kernel of the integral operator
\begin{eqnarray}
\hspace{-1cm}K_{\mu\alpha}(\mathbf{r},\mathbf{r}'';\omega)&=&-D^{(0)}_{\mu\alpha}(\mathbf{r}-\mathbf{r}'';\omega)\chi(\mathbf{r}'')%
\nonumber\\%
&&-\sum_{s}E^{(s)}_{\mu}(\mathbf{r})\,D^{(s)\ast}_{\alpha}(\mathbf{r}'')%
\label{a.13}
\end{eqnarray}
The sign of "approximately equal" in Eq.~(\ref{a.12}) means that here we have equated the waveguide modes with the eigenfunctions of the original integral operator (\ref{a.3}) and have neglected the differences in the eigenvalues such that $\Lambda_k\approx 1$. In this assumption the integral operator with kernel (\ref{a.13}) eliminates the waveguide contribution (\ref{a.10}) and, as a consequence, justifies replacing $D^{(E)}\to D^{(\mathrm{ext})}$ in the right-hand side of Eq.~(\ref{a.12}).

The inhomogeneous integral equation (\ref{a.12}) fulfills the resolving conditions of the Fredholm theorem, see Ref.~\cite{KurantHilbert}, and its solution for a nanofiber implies rapidly converging iterative expansion. In the first iteration step we have
\begin{eqnarray}
\lefteqn{\hspace{-1cm}D^{(\mathrm{ext})}_{\mu\nu}(\mathbf{r},\mathbf{r}';\omega)\approx \tilde{D}^{(0)}_{\mu\nu}(\mathbf{r},\mathbf{r}';\omega)}%
\nonumber\\%
&&\hspace{-1cm}-\frac{1}{\hbar}\int\!d^{3}r''\,D^{(0)}_{\mu\alpha}(\mathbf{r}-\mathbf{r}'';\omega)\chi(\mathbf{r}'')D^{(0)}_{\alpha\nu}(\mathbf{r}''-\mathbf{r}';\omega)%
\label{a.14}
\end{eqnarray}
where the first term is given by
\begin{eqnarray}
\lefteqn{\hspace{-0.7cm}\tilde{D}^{(0)}_{\mu\nu}(\mathbf{r},\mathbf{r}';\omega)=D^{(0)}_{\mu\nu}(\mathbf{r}-\mathbf{r}';\omega)}%
\nonumber\\%
&&\hspace{-1cm}-\sum_{s}E^{(s)}_{\mu}(\mathbf{r})\!\int\!\!d^3r''\,D^{(s)\ast}_{\alpha}(\mathbf{r}'')D^{(0)}_{\alpha\nu}(\mathbf{r}''-\mathbf{r}';\omega)%
\label{a.15}
\end{eqnarray}
and coincides with the expansion (\ref{a.11}) with the vacuum Green's function substituted in the right-hand side. The second term in Eq.~(\ref{a.14}) gives the contribution of a single scattering from a spatial inhomogeneity in the dielectric permittivity i.e. gives the simplest perturbative estimate for the reflection from the waveguide.

The solution of the scattering equation in the integral form (\ref{a.12}) suggests a realistic estimate of the function $D^{(\mathrm{ext})}$ in the considered assumptions and simplifications. As a zero approximation we can accept
\begin{equation}
D^{(\mathrm{ext})}_{\mu\nu}(\mathbf{r},\mathbf{r}';\omega)\approx\tilde{D}^{(0)}_{\mu\nu}(\mathbf{r},\mathbf{r}';\omega)%
\label{a.16}
\end{equation}
whose validity is explained below. For the propagating modes the evanescent field typically has transverse scale sufficiently broader than the radiation wavelength. Then the second term in Eq.~(\ref{a.15}) can be approximately evaluated for points $\mathbf{r}$ and $\mathbf{r}'$ separated by a distance comparable with the wavelength or shorter and the complete Green's function (\ref{a.9}) can be given in the following closed form
\begin{eqnarray}
\lefteqn{\hspace{-0.7cm}D^{(E)}_{\mu\nu}(\mathbf{r},\mathbf{r}';\omega)\approx D^{(\mathrm{wg})}_{\mu\nu}(\mathbf{r},\mathbf{r}';\omega)+D^{(0)}_{\mu\nu}(\mathbf{r}\!-\!\mathbf{r}';\omega)}%
\nonumber\\%
&&\hspace{-0.7cm}-\sum_{s}\frac{4\pi\hbar\omega^2}{\omega^2-c^{2}k^{2}+i0}\,E^{(s)}_{\mu}(\mathbf{r})\,E^{(s)\ast}_{\bot\nu}(\mathbf{r}')%
\label{a.17}
\end{eqnarray}
where in the last subtracting term $\mathbf{E}^{(s)\ast}_{\bot}(\mathbf{r}')$ denotes the vector projection of $\mathbf{E}^{(s)\ast}(\mathbf{r}')$ on the plane transverse to $z$-axis i.e. to the waveguide direction, see Eq.~(\ref{3.2}). It is given by the leading (in paraxial limit) contributions in the right-hand side of (\ref{3.2}) with omitted the azimuthal angular dependent terms.

The obtained result has a quite natural physical interpretation. If a point-like dipole source emits light near a single-mode waveguide of sub-wavelength transverse scale then the respective fundamental solution of the Maxwell equation has more or less a similar structure as in vacuum case. Then, part of the emitted modes in the paraxial approach coincide with the waveguide modes, but in the presence of the waveguide these modes have the dispersion law $\omega_s=\omega_k\neq c k$ different from the dispersion relation in free space and this difference should be taken into consideration. The respective correction can be seen by comparing the waveguide contribution (\ref{a.10}) with the last term in the expression for the complete Green's function (\ref{a.17}). In the paraxial approach the subtracted contribution, given by this last term, eliminates the emission into the vacuum mode coinciding, in paraxial limit, with the waveguide modes, whose contribution is already correctly incorporated into Eq.~(\ref{a.10}). The approximation, performed by Eqs.~(\ref{a.16}) and (\ref{a.17}), seems sufficient for a dipole source distant from the fiber on a length comparable with its transverse scale. Nevertheless, it becomes insufficient for a dipole located just near the fiber surface since it completely ignores any corrections associated with the reflection of the source wave from the fiber. The respective corrections could be recovered via an iterative solution of Eq.~(\ref{a.12}) and their simplest estimate is given by the second term in Eq.~(\ref{a.14}).

\section{The waveguide modes} \label{Appendix_B}

\noindent We consider the waveguide mode equations in the cylindrical frame. By substituting Eq.~(\ref{a.6}) into Eq.~(\ref{a.5}) for the longitudinal $z$-component for both types of $\sigma=\pm 1$ polarization modes we have
\begin{equation}
\frac{\partial^2}{\partial\rho^2}U^{(\sigma k)}(\rho) +\frac{1}{\rho}\frac{\partial}{\partial\rho}U^{(\sigma k)}(\rho)+\left[\kappa^2-\frac{\sigma^2}{\rho^2}\right]U^{(\sigma k)}(\rho)=0%
\label{b.1}
\end{equation}
where $\kappa^2=\kappa_{\mathrm{in}}^2=\epsilon\,\omega^2/c^2-k^2>0$ (inside the fiber $\rho<a$) or $\kappa^2=\kappa_{\mathrm{out}}^2=\omega^2/c^2-k^2<0$ (outside the fiber $\rho>a$) and $U^{(\sigma k)}(\rho)$ is either electric field $E_z^{(\sigma k)}(\rho)$ or magnetic field component $H_z^{(\sigma k)}(\rho)$ (expanded similarly to (\ref{a.6})). Other vector components are given by
\begin{eqnarray}
E_\rho^{(\sigma k)}(\rho)&=&\frac{i}{\kappa^2}\left[k\frac{\partial}{\partial\rho}E_z^{(\sigma k)}(\rho)+\frac{i\sigma\omega}{c\rho}H_z^{(\sigma k)}(\rho)\right]%
\nonumber\\%
\nonumber\\%
E_\phi^{(\sigma k)}(\rho)&=&\frac{i}{\kappa^2}\left[\frac{i\sigma k}{\rho}E_z^{(\sigma k)}(\rho)-\frac{\omega}{c}\frac{\partial}{\partial\rho}H_z^{(\sigma k)}(\rho)\right]%
\nonumber\\%
\label{b.2}%
\end{eqnarray}
for electric field and
\begin{eqnarray}
H_\rho^{(\sigma k)}(\rho)&=&\frac{i}{\kappa^2}\left[k\frac{\partial}{\partial\rho}H_z^{(\sigma k)}(\rho)-\frac{i\sigma\epsilon\omega}{c\rho}E_z^{(\sigma k)}(\rho)\right]%
\nonumber\\%
H_\phi^{(\sigma k)}(\rho)&=&\frac{i}{\kappa^2}\left[\frac{i\sigma k}{\rho}H_z^{(\sigma k)}(\rho)+\frac{\epsilon\omega}{c}\frac{\partial}{\partial\rho}E_z^{(\sigma k)}(\rho)\right]%
\nonumber\\%
\label{b.3}
\end{eqnarray}
for magnetic field. These relations are written for arbitrary $\rho$ such that the dielectric constant should be taken as $\epsilon=1$ outside the fiber. The mode equation (\ref{b.1}), considered together with representations of transverse components (\ref{b.2}) and (\ref{b.3}), has to be completed by conventional boundary conditions to the Maxwell equations on the fiber surface. Eventually the solution can be found in an analytical form as compilation of Bessel functions.

We set the basic functions $E_{\rho}(\rho)$, $E_{\phi}(\rho)$, and $E_{z}(\rho)$ as the electric field components for the right-handed rotating polarization mode with $\sigma=+1$, see Eq.~(\ref{3.1}) in the main text. These functions can be expressed via the following expansion in the set of the Bessel and Hankel functions of the first kind, see \cite{Marcuse}
\begin{eqnarray}
E_{\rho}(\rho)&\propto&\frac{i\,k}{2\kappa\,J_1(\kappa a)}\left[(1-u)J_0(\kappa\rho)-(1+u)J_2(\kappa\rho)\right]%
\nonumber\\%
E_{\phi}(\rho)&\propto&-\frac{k}{2\kappa\,J_1(\kappa a)}\left[(1-u)J_0(\kappa\rho)+(1+u)J_2(\kappa\rho)\right]%
\nonumber\\%
E_{z}(\rho)&\propto&\frac{1}{J_1(\kappa a)}\,J_1(\kappa \rho)%
\label{b.4}
\end{eqnarray}
with $\kappa=\kappa_{\mathrm{in}}$ as real quantity for $\rho<a$ and
\begin{eqnarray}
E_{\rho}(\rho)\!&\propto&\!-\frac{i\,k}{2\kappa\,H_{1}^{(1)}\!(\kappa a)}\left[(1-u)H_0^{(1)}\!(\kappa\rho)-(1+u)H_2^{(1)}\!(\kappa\rho)\right]%
\nonumber\\%
E_{\phi}(\rho)\!&\propto&\!\frac{\,k}{2\kappa\,H_{1}^{(1)}\!(\kappa a)}\left[(1-u)H_0^{(1)}\!(\kappa\rho)+(1+u)H_2^{(1)}\!(\kappa\rho)\right]%
\nonumber\\%
E_{z}(\rho)\!&\propto&\!\frac{1}{H_{1}^{(1)}\!(\kappa a)}\,H_{1}^{(1)}\!(\kappa \rho)%
\label{b.5}
\end{eqnarray}
with $\kappa=\kappa_{\mathrm{out}}$ as imaginary quantity for $\rho>a$, where
\begin{eqnarray}
\lefteqn{u=-\frac{\omega^2(\epsilon-1)}{c^2\,a^2\kappa_{\mathrm{in}}^2\kappa_{\mathrm{out}}^2}}%
\nonumber\\%
&&\hspace{-0.5cm}\times\!\!\left[ \left.\frac{1}{x}\frac{d}{dx}\ln J_1(x)\right|_{x=\kappa_{\mathrm{in}}a}\!-\!\left.\frac{1}{x}\frac{d}{dx}\ln H_1^{(1)}(x)\right|_{x=\kappa_{\mathrm{out}}a}\right]^{-1}%
\label{b.6}%
\end{eqnarray}
The mode functions should be normalized in accordance with Eq.~(\ref{a.8}) and the modes obey the dispersion law $k=k(\omega)$, which is given by the solution of the following characteristic equation
\begin{eqnarray}
\lefteqn{\left[ -\epsilon\frac{\kappa_{\mathrm{out}}^2}{\kappa_{\mathrm{in}}^2}\left.x\frac{d}{dx}\ln J_1(x)\right|_{x=\kappa_{\mathrm{in}}a}\!%
+\!\left.x\frac{d}{dx}\ln H_1^{(1)}(x)\right|_{x=\kappa_{\mathrm{out}}a}\right]}%
\nonumber\\%
&&\times\left[-\frac{\kappa_{\mathrm{out}}^2}{\kappa_{\mathrm{in}}^2}\left.x\frac{d}{dx}\ln J_1(x)\right|_{x=\kappa_{\mathrm{in}}a}\!%
+\!\left.x\frac{d}{dx}\ln H_1^{(1)}(x)\right|_{x=\kappa_{\mathrm{out}}a}\right]%
\nonumber\\%
\nonumber\\%
&&=\frac{(\epsilon-1)^2\omega^2 k^2}{c^2\kappa_{\mathrm{in}}^4}
\end{eqnarray}

\end{document}